\documentstyle[12pt]{article}
\begin{document}

\centerline{\bf \large Quantum Computation in Quantum-Hall Systems}

\ 

\centerline{V.~Privman,$^{1}$ I.~D.~Vagner$^2$ and G.~Kventsel$^{2,3}$}

\ 

\ 

\noindent\hang{$^{1}$Department of Physics, Clarkson University, 
Potsdam,\hfill\break{}New York 13699--5820, USA}

\noindent\hang{$^{2}$Grenoble High Magnetic Field Laboratory,
Max-Planck-Institut f\"ur\hfill\break{}Festk\"orperforschung, and
Centre National de la Recherche\hfill\break{}Scientifique,
BP 166, F-38042, Grenoble Cedex 9, France}

\noindent\hang{$^{3}$Department of Chemistry, Technion---Israel Institute of
Technology,\hfill\break{}Haifa 32000, Israel}

\ 

\ 

\ 

\begin{abstract}
We describe a quantum information processor (quantum computer) based on the
hyperfine interactions between the conduction electrons and nuclear spins
embedded in a two-dimensional electron system in the quantum-Hall regime.
Nuclear spins can be controlled individually by
electromagnetic pulses. Their interactions, which are of the spin-exchange type, can be possibly switched on and off pair-wise dynamically, for
nearest neighbors, by controlling impurities. We also propose the way to
feed in the initial data and explore ideas for reading off the final
results.
\end{abstract}

\vfill{}\centerline{\tt Submitted to Physics Letters A}

\newpage

The field of quantum computing has seen an explosive growth of theoretical
development [1-7]. It has been realized that
quantum computers can be faster than classical computers for some problems 
[1-3,8-13]. The analog nature of errors and
possible error correction schemes have been explored [6,7,9,13-21].
There have also been several proposals for
actual realizations of quantum information processing [4,5,13,22-31]. 
Two of these proposals: the
ion-trap system [5,22,25,27,28] and the ensemble-of-molecules
liquid-state NMR approach [29-31] have been studied extensively
as possible experimental realizations of quantum computing.
However, all experimental results to date only accomplish
the simplest quantum-logic functions such as single-spin rotations
or two-spin controlled-NOT [1-7].

A major challenge faced by both experiment and theory has involved scaling up 
from one to many quantum gates and actual ``programming,'' i.e., conducting
calculations by coherent quantum unitary evolution, in a controlled fashion.
Experimentally, quantum computation requires switching on and off
pair-wise interactions between various two-state systems, e.g., spins ${1 \over 2}$,
termed ``qubits.'' 
Initialization and reading
off the final results are also nontrivial parts of the process. 
Ideally, the latter should involve efficient measurement of
a single qubit. The NMR variant [29-31] measures instead ensemble
averages (expectation values). Certain ``fault-tolerant'' error 
correction schemes [7,13,17,19-21]
actually also require measurements of some of
the qubits {\it during\/} the computation.

Theoretically, the most striking recent development has been the formulation
of the fault-tolerant error correction schemes [7,13,17,19-21]. Correction of
analog errors inherent in quantum computation due to the
superposition-of-states property (which in turn is central to the speed-up
of some calculations) means an uphill battle against the second law of
thermodynamics. These error-correction schemes [7,13,17,19-21] aim at
calculations that can go on indefinitely provided the overall error rate 
at each qubit is small enough.

It is not our goal here to review these issues: we will adopt the
point of view that modern error correction schemes will allow
calculations long enough to be useful provided a working quantum information
processor can be devised. It is the latter aspect that we address in this
work. Thus, we propose a quantum computer realization based on hyperfine
interactions [32] between the conduction electrons and nuclear
spins embedded in a two-dimensional electron system in 
which the electron gas is in the quantum-Hall
effect (QHE) regime [33,34]. Such systems have been made at the
interfaces between semiconductor materials and in superlattices (layered
semiconductor structures) [35].

In these systems, at temperatures of order 1K and applied magnetic fields of
several Tesla, there are intervals of magnetic field values for which
the electrons fill up an integer number of Landau levels [36]. 
The electron gas then forms a nondissipative QHE fluid [35]; the Hall
resistance exhibits a plateau at a value that is a multiple of $e^2/h$,
while the dissipativity of the conduction electron gas (the
magnetoresistance) approaches zero. Nuclear-spin thermalization/relaxation
processes occur on the time scale denoted $T_1$ [32] which,
experimentally, ranges from several minutes to half an hour
[37-39]. It is
expected then that the nuclear spin dynamics is dominated by coherent spin
exchanges mediated by electrons [40,41].
Owing to rapid advances in the experimental facilities, the hyperfine
interactions in QHE systems have recently attracted growing theoretical 
[41,42] and experimental [37-39] interest; this progress makes it 
feasible to handle the electron spin-nuclear spin
interactions with almost atomic precision.

Similar to the ion-trap system [5,22,25,27,28], we consider a
chain of spin-${1 \over 2}$ nuclei, of atoms positioned by the molecular-beam epitaxy 
techniques [35] in an effectively two-dimensional
system subjected to a strong magnetic field. The
typical separation should be comparable to the magnetic length $\ell_H=
\sqrt{\hbar c/eH}$, where $H$ is the applied magnetic field, perpendicular
to the two-dimensional layer. This length is of the order of $100\,$\AA{}. We
propose to control individual nuclear spins by electromagnetic-radiation
pulses in the nuclear magnetic resonance (NMR) frequency range [32]. 

An important question is how to control nuclear spins individually.
Use of magnetic field gradient could be 
contemplated to achieve differentiation, but there are severe 
limitations on the field variation owing
to the need to maintain the QHE electronic state.
Instead, one can use {\it different nuclei}.
Theoretically, there is no apparent limit
on how many different spins can be arranged in a chain. However,
practically the number of suitable spin-${1\over 2}$ isotopes may be
limited. Thus, achieving sufficient chemical-shift dispersion for
systems of more than few qubits may require additional ideas;
the following ones are tentative because presently it is not known
how realistic these proposals are from the point of view of actual
experimental realizations. Specifically, one can position nuclear spins
in different crystalline environments. The
latter can be controlled by implanting atoms and complexes into the host
material [32]. It may be also possible 
to utilize small clusters of nuclear spins, rather than
individual spins. These can be made coherent [43] by lowering the temperature to
order several $\mu $K, as compared to order 1K needed to achieve the QHE state.

Under the typical conditions of QHE the direct dipole-dipole interaction of
the nuclear spins is negligibly small [41]. The dominant interaction
will be that mediated by the contact hyperfine interactions between nuclear
spins and conduction electrons [40]. Similarly, electron-mediated
interactions leading to the scalar coupling have been utilized in the
liquid-state NMR realization of quantum computation [29-31]. In ordinary
metals, the electron-mediated nuclear-spin interactions 
exhibit Friedel oscillations [32] because of
the existence of sharp Fermi surface.

In the quantum-Hall regime, however, the energy spectrum of the
two-dimensional electron gas is discretized by the magnetic field. As a
result, the interaction is no longer oscillatory but rather monotonic,
exponentially decaying [40] on the length scale $\ell_H$. The
following terms in the effective phenomenological two-spin interaction
Hamiltonian correspond to the second-order perturbative
calculation (carried out for two identical nuclei) of [40], where
for different nuclei we replaced $Z^2$ by the product of the two atomic
numbers (which is basically a guess),

\begin{equation}
-VZ^{(1)}Z^{(2)}H^{-1}\sqrt{\frac{c\ell _H}r}e^{-cr/\ell _H}\left[ \sigma
_{-}^{(1)}\sigma _{+}^{(2)}+\sigma _{+}^{(1)}\sigma _{-}^{(2)}\right] \,,
\end{equation}

\noindent where $c$ is a dimensionless quantity [40] of order 1, $Z^{(j)}$
are the atomic numbers of the nuclei, while $V$ is some constant.
Note that

\begin{equation}
r/\ell _H\propto r\sqrt{H}\,.
\end{equation}

\noindent Here $H$ is the applied field, $r$ is the spin-spin separation,
while $\sigma^{(j)}$ are the Pauli matrices corresponding to the spin-${1 \over 2}$
operators of the two nuclei labeled by the superscripts $j=1,2$. Each
nuclear spin also interacts with the applied field via the magnetic coupling
of the form $-\gamma^{(j)}\hbar H \sigma_z^{(j)}$. Determination of the
precise effective
spin-spin interaction Hamiltonian will likely to be accomplished to a
large extent by direct experimental probe. The strength of the interaction
in Eq.~(1) can be roughly estimated to be of order $10^{-16}\,$erg,
which corresponds to frequency of order $10^{11}\,$Hz.

For quantum computation, one has to devise the means to control the spin-spin
interactions. Ideally, one would like to be able to switch interactions on
and off at will, for varying time intervals $\Delta t$. Switching on a
pair-wise interaction would allow to carry out a unitary transformation on a
pair of spins independently of the other spins. It has been established
[13,23,44-47] that nearly any such transformation, combined with
single-spin transformations which can be accomplished by radiation pulses,
form a universal set in the sense that arbitrary ``computer program'' can be
built from them. There are NMR ``refocusing'' 
methods that allow such control, as utilized, 
for instance, in the liquid-state NMR formulation [29-31]
of quantum computing. However, until the full form of the spin-spin
interaction Hamiltonian is established for our case, it is useful to 
consider other ideas as well.

Geometry constraints would limit the pairs of spins for which the two-spin
interactions are nonnegligible typically to nearest-neighbor pairs.
Furthermore, other interactions cannot be really {\it fully\/} eliminated,
but only reduced. Still, control of the spin-spin interactions would allow
added flexibility in ``programming'' the unitary evolution of a
computational device. Even when the control is possible, in practice it
would be unrealistic to expect the form of the interaction, such as Eq.~(1)
above, be known exactly from theoretical calculations alone. Thus, Eq.~(1)
is a leading-approximation/guess phenomenological form. Input from
experiments will
be required to fine-tune the computer functions that depend on such internal
interactions.

One possibility not based on the NMR methods is
to disrupt (ideally, switch off), for the duration of some
time interval $\Delta t$, the interaction for one (nearest-neighbor)
pair of spins by placing impurities
between the spins, see Figure~1. The impurities can be ionized by external
electromagnetic pulses to electronic configurations that capture electrons
and locally destroy the coherence of the electron gas. Differentiation can
be achieved by using different impurity species. Admittedly, this is a rather
speculative idea. Specifically, it may be more appropriate to place
the impurities near or surrounding the nuclear spins, instead of the geometry of
Figure~1.

It is important to emphasize that the pair-wise interactions are ``on'' most
of the time, for each pair of spins. Therefore, the ``idle'' unitary
transformations in the latter approach will not be simple phase 
changes as for noninteracting
spins. The ability to change the interactions locally, pair-wise, will only 
allow to change the {\it relative\/} unitary transformations to which
nearest-neighbor spin pairs are subject. In addition, one has the
single-spin rotations that can be done by external electromagnetic pulses.
Programming of such a computer is therefore less straightforward than
usually expected in the theoretical approaches that 
assume {\it noninteracting\/} idling elements [1-7,13,23,44-47]; however, this
is only a matter of new mathematical
developments being called for.

We now turn to the process of ``feeding in'' the initial data into the
computer. This can be accomplished as follows: initially, all
the nuclear spins in the system are pumped in one direction. This can be
achieved by shining a polarized light at the system [49] that
creates electron-hole pairs. These pairs annihilate, forcing on a fixed
nuclear spin polarization, corresponding to that of the incident light [49].
After the initial alignment, the nuclear spins can be rotated to the desired
quantum states needed for computation by electromagnetic pulses at their
respective frequencies.

In all the proposals for quantum computation [1-7,13,22-31],
reading off the final spin states by
measuring, and also the measurement processes that are required for error
correction [7,13,17,19-21], are most challenging to realize. This is 
because direct interaction of a microscopic system with any macroscopic
system for the purpose of measurement is disruptive and difficult to carry
out in an orderly fashion for all the individual spins in the system.
 
We note that as for the liquid-NMR proposal [29-31], we could read off
averages by NMR techniques by producing replicas of the spin chain, 
see Figure~1, and letting
them evolve in parallel. 
The electromagnetic pulses that control the
computation can be applied to all the replicas at once. However, some quantum
error correction protocols [7,13,17,19-21] require actual measurements rather
than averages. Furthermore, unlike the liquid-state NMR, there may be uncontrollable
differences between the replicas. The only thing that might save the situation is
the fact that our spins are located at distances much larger than atomic dimensions.
Therefore, some averaging of the ``atomic'' scale influences may be expected in the
spin-spin interactions controlling the actual computation in each chain. The latter
observation suggests that measurement methods other than NMR-based must be
explored. We propose three measuring processes below: the first and second
may be more appropriate for 
final-state readout while the second and third for error-correction schemes.

First, let us assume that the final state is one of the 
direct-product states of the $n$-spin system. It is possible to generate by
holographic and other methods [48-51] a narrow
strip of conductance at each spin in turn, see
Figure~1, and send a current of spin-polarized electrons through it. The
observed current can be pre-calibrated to enable high certainty
determination of whether there was a spin-exchange scattering event thus
determining the nuclear spin's direction, resembling the spin-diode
[38,52] techniques. Furthermore, one can extend the strip of conductance
over several replicas of
the spin chain, separated order of magnitude more than the spins, e.g.,
$1000\,$\AA{}. One can probably have enough of them to reduce
significantly any uncertainty in the spin direction determination.

Second, if the final or intermediate state (the latter case is relevant 
for error correction) can be 
entangled, so that one cannot simply measure
each spin in turn, then the situation is more complicated. One can
generate a ``mask'' of conducting strips, for all or a group of spins.
However, ``calibration'' to derive data pertinent to the multispin
quantum state may be a challenge.

Third, some error correction schemes [7,13,17,19-21] require 
measurement of difference of
the components of nearby spins. This might be contemplated by having two
conducting strips with the spin-polarized electron current, and adding a
time-dependent component to the applied magnetic field for the duration of
the measurement. Difference in the nuclear spin states will then affect the
Aharonov-Bohm oscillatory structure of the observed current; see [53] for
survey of such effects.

In summary, we have proposed a model of a quantum
computer based on the hyperfine interactions between the electron and nuclear
spins in quantum Hall effect systems. This brings to two the number 
of proposals that have been formulated theoretically for realizations 
of quantum computing which can be potentially done in 
{\it solid-state systems}; the other is the quantum-dot proposal [26].
The possibility of quantum computing in solid-state is exciting. Indeed, the intricacies of modern technology, especially as far as nanoscale ``engineering'' is concerned, are much more geared for solid-state systems than any other medium. All modern electronic devices, with, presently, components on submicron scales, are solid-state.

However, unlike the more ``established'' quantum computing proposals such as ion traps and liquid-state NMR, the two solid-state proposals are presently theoretical. There are several investigations needed, of the form and strength of the spin-spin interactions, of the time scales of interaction vs. decoherence, and other topics, before initial experimental attempts to build few-qubit QHE quantum-computing systems can be deemed realistic. Specifically, no estimates are available of the time scales of decoherence which may be orders of magnitude shorter than $T_1$.

We wish to thank D.~Mozyrsky for helpful comments on the manuscript
and P.~Wyder for the hospitality at
Grenoble HFML and interest in this work.
The work of V.P. has been supported in part by US Air
Force grants, contract numbers F30602-96-1-0276 and F30602-97-2-0089. 
I.V. acknowledges support by a grant from the German-Israeli
Foundation for Scientific Research and Development, number
G 0456-220.07/95.

\newpage

\centerline{\bf FIGURE CAPTION}

\ 

\noindent\hang {\bf Figure 1:}\ \ \ The schematics of the proposed two-dimensional
nuclear-spin system: N denotes atoms with spin-$1\over 2$ nuclei; I denotes impurity atoms
or complexes that can be ionized to disrupt the spin-exchange interactions mediated
by conduction electrons (the impurity placement may be different, see text); R illustrate replicas (actually there will be many of them);
E and C represent conducting electrodes and connecting strip for measurement
(see text).

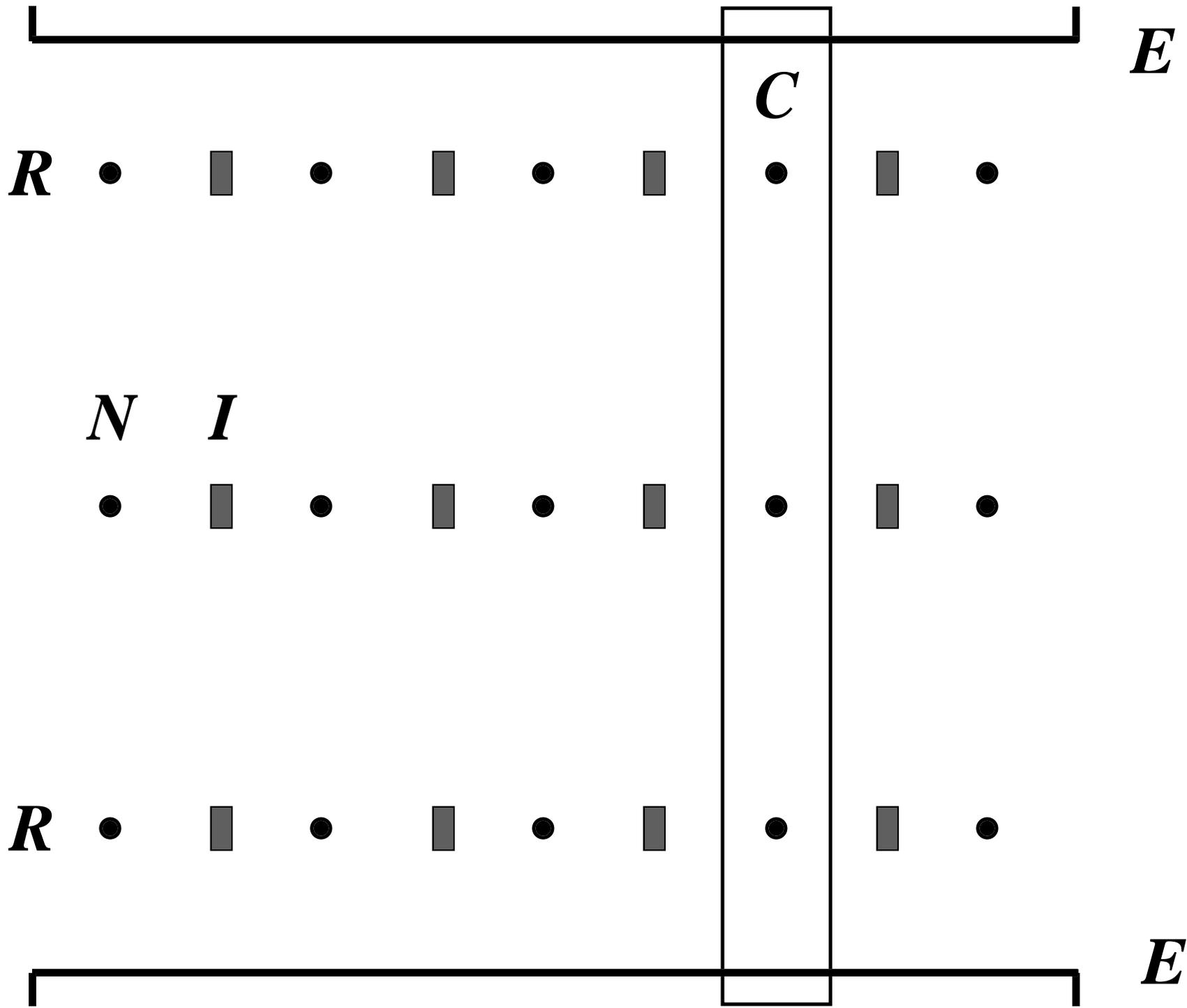
\end{document}